\documentstyle[amssymb,11pt]{article}

\input{tcilatex}

\begin{document}

\small

\title{Finite-size scaling of the error threshold transition in finite populations}
\author{P. R. A. Campos and J. F. Fontanari \\
Instituto de F\'{\i}sica de S\~ao Carlos \\
Universidade de S\~ao Paulo \\
Caixa Postal 369 \\
13560-970 S\~ao Carlos SP \\
Brazil}
\date{}
\maketitle

\centerline{\large{\bf Abstract}}

\bigskip

The error threshold transition in a stochastic (i.e. finite population)
version of the quasispecies model of molecular evolution is studied using
finite-size scaling. For the single-sharp-peak replication landscape, the
deterministic model exhibits a first-order transition at $Q=Q_c=1/a$, where $%
Q$ is the probability of exact replication of a molecule of length $L
\rightarrow \infty$, and $a$ is the selective advantage of the master
string. For sufficiently large population size, $N$, we show that in the
critical region the characteristic time for the vanishing of the master
strings from the population is described very well by the scaling assumption 
$\tau = N^{1/2} f_a \left [ \left ( Q - Q_c \right ) N^{1/2} \right ] $,
where $f_a$ is an $a$-dependent scaling function.

\bigskip

\bigskip

{\bf Short Title:} error threshold in finite populations


\newpage


An elusive issue in the extension of Eigen's quasispecies model \cite{Eigen}
of molecular evolution to finite populations is the characterization of the
so-called error threshold phenomenon which limits the length of the
molecules and, consequently, the amount of information they can store \cite
{reviews}. This phenomenon poses an interesting challenge to the theories of
the origin of life, since it prevents the emergence of huge molecules which
could carry the necessary information for building a complex metabolism.
Moreover, since modern theories of integration of information in pre-biotic
systems involve the compartmentation of a small number of molecules, the
understanding of the effects of the error propagation in finite populations
has become a major issue to the theories of the origin of life \cite{book}.

The quasispecies model was originally formulated within a deterministic
chemical kinetic framework based on a set of ordinary differential equations
for the concentrations of the different types of molecules that compose the
population. Such formulation, however, is valid only in the limit where the
total number of molecules, denoted by $N$, goes to infinity. In the binary
version of the quasispecies model, a molecule is represented by a string of $%
L$ digits $\left (s_1, s_2, \ldots, s_L \right )$, with the variables $%
s_\alpha$ allowed to take on only two different values, say $s_\alpha = 0,1$%
, each of which representing a different type of monomer used to build the
molecule. The concentrations $x_i$ of molecules of type $i =1, 2, \ldots,
2^L $ evolve in time according to the equations \cite{Eigen,reviews} 
\begin{equation}  \label{ODE}
\frac{dx_i}{dt} = \sum_j W_{ij} x_j - \Phi \left ( t \right ) x_i \; ,
\end{equation}
where $\Phi (t)$ is a dilution flux that keeps the total concentration
constant. This flux introduces a nonlinearity in Eq.\ (\ref{ODE}), and is
determined by the condition $\sum_i dx_i /dt = 0$. In particular, assuming $%
\sum_i x_i = 1$ yields 
\begin{equation}  \label{flux}
\Phi = \sum_{i,j}W_{ij} x_j \; .
\end{equation}
The elements of the replication matrix $W_{ij}$ depend on the replication
rate or fitness $A_i$ of the strings of type $i$ as well as on the Hamming
distance $d \left ( i,j \right )$ between strings $i$ and $j$. They are
given by \cite{Eigen,reviews} 
\begin{equation}
W_{ii} = A_i \, q^\nu
\end{equation}
and 
\begin{equation}
W_{ij} = A_j \, q^{L - d \left ( i,j \right )} \left ( 1 - q \right )^{d
\left ( i,j \right )} ~~~~i \neq j ,
\end{equation}
where $0 \leq q \leq 1$ is the single-digit replication accuracy, which is
assumed to be the same for all digits.

The quasispecies concept and the error threshold phenomenon are illustrated
more neatly for the single-sharp-peak replication landscape, in which we
ascribe the replication rate $a > 1$ to the so-called master string, say $%
\left (1, 1, \ldots, 1 \right )$, and the replication rate $1$ to the
remaining strings. In this context, the parameter $a$ is termed selective
advantage of the master string. As the replication accuracy $q$ decreases,
two distinct regimes are observed in the population composition in the
deterministic case: the {\it quasispecies} regime characterized by the
presence of the master string together with its close neighbors, and the 
{\it uniform} regime where the $2^L$ strings appear in the same proportion.
The transition between these regimes takes place at the error threshold $q_c$%
, whose value depends on the parameters $L$ and $a$ \cite{Eigen,reviews}.
However, even in the deterministic case, $N \rightarrow \infty$, a genuine
thermodynamic order-disorder phase transition will occur in the limit $L
\rightarrow \infty$ only \cite{Lethausser,Tarazona,Gal}. To study this
transition for large $L$, it is convenient to introduce the probability of
exact replication of an entire string, namely, 
\begin{equation}
Q = q^L \; ,
\end{equation}
so that the {\it discontinuous} transition occurs at 
\begin{equation}  \label{Q_c}
Q_c = \frac{1}{a} \;
\end{equation}
for $L \rightarrow \infty$ \cite{Eigen,Gal}. A recent finite-size scaling
study of the sharpness of the threshold transition indicates that the
characteristics of the transition persist across a range of $Q$ of order $%
L^{-1}$ about $Q_c$ \cite{Paulo}.

Although several theoretical frameworks have been proposed to generalize the
deterministic kinetic formulation of the quasispecies model so as to take
into account the effect of finite $N$ \cite
{Ebeling,McCaskill,Nowak,Yi,Bonnaz,Alves}, the somewhat uncontrolled
approximations used in those analyses have hindered a precise
characterization of the error threshold for finite populations. In
particular, Nowak and Schuster \cite{Nowak} employed a simple birth and
death model, whose deterministic limit, however, does not yield the
stationary distribution predicted by Eq. (\ref{ODE}), as well as numerical
simulations based on Gillespie's algorithm \cite{Gillespie} to show that an
appropriately defined $Q_c (N)$ tends to the deterministic value $1/a$ with $%
N^{-1/2}$ for sufficiently large populations. A similar result was obtained
by neglecting the possibility of occurrence of multiple errors during the
replication of a molecule \cite{Bonnaz}. A more drastic approximation that
neglects linkage disequilibrium at the population level yields that $Q_c(N)$
increases linearly with $1/N$ \cite{Alves}. Of course, since there is no
generally accepted definition of error threshold for finite $N$ (and for
finite $L$ as well), denoted above by $Q_c(N)$, there are some arbitrariness
in those analyses.

In this paper we follow a more direct approach to characterize the error
threshold transition for finite $N$, which dispenses with a definition for $%
Q_c(N)$. As mentioned before, since a genuine phase transition occurs in the
limits $N \rightarrow \infty$ and $L \rightarrow \infty$ only, we study a
stochastic (i.e. finite $N$) version of the quasispecies model with $L
\rightarrow \infty$ and $q \rightarrow 1$ so that $Q = q^L $ is finite. In
this limit the problem simplifies enormously as the probability of any
string becoming a master string due to replication errors is of order $1/L$
and so can be safely neglected. Besides, since for the single-sharp-peak
replication landscape the strings can be classified in two types only: the
master strings and the error tail, which comprises all other strings, the
population at any given generation can be described by the single integer $n
= 0, 1, \ldots, N$, which gives the number of master strings in the
population. The goal then is to calculate the probability distribution that
at generation $t$ there are exactly $n$ master strings in the population.
This quantity, denoted by ${\cal {P}}_t (n)$, obeys the recursion equation 
\begin{equation}  \label{recursion}
{\cal {P}}_{t+1} (n) = \sum_{m=0}^N T(n,m) \, {\cal {P}}_t (m)
\end{equation}
with the elements of the transition matrix ${\bf T}$ given by 
\begin{equation}  \label{transition}
T (n,m) = \sum_{k=n}^m \, \left ( \! \! 
\begin{array}{c}
N \\ 
k
\end{array}
\! \! \right ) \, \left ( \! \! 
\begin{array}{c}
k \\ 
n
\end{array}
\! \! \right ) \, w_m^k \left ( 1 - w_m \right )^{N-k} \, Q^n \left (1 - Q
\right )^{k -n} \; ,
\end{equation}
where 
\begin{equation}  \label{w}
w_m = \frac{ m a}{N - m + m a}
\end{equation}
is the relative fitness of the master strings. In writing Eq.\ (\ref
{transition}) we have followed the prescription used in the implementation
of the standard genetic algorithm \cite{Goldberg}: first the natural
selection process acting via differential reproduction is considered and
then the mutation process. We note that $\sum_n T(n,m) = 1 ~\forall m$ and $%
T(0,0) = 1$. Moreover the largest eigenvalue of ${\bf T}$ is $\lambda_0 = 1$
and its corresponding eigenvector is ${\bf l}_0^\dagger = (1,0,\ldots,0)$.
This stochastic model is easily recognized as the celebrated Kimura-Crow
infinite allele model \cite{KC,Kimura} which has been extensively studied
within the diffusion approximation for large $N$. However, for arbitrary
values of $Q$ and $a$ the solutions of the partial differential equations
are too complicated to be of any utility to our purposes \cite{Kimura}.

As for finite $N$ the fluctuations, either in the reproduction or mutation
processes, will ultimately lead to the irreversible loss of all copies of
the master string from the population, the asymptotic solution of Eq.\ (\ref
{recursion}) is simply ${\cal {P}}_\infty (n) = \delta_{n0}$. Our goal is to
determine how the characteristic time, $\tau$, that governs the vanishing of
the master strings from the population depends on $N$, $Q$ and $a$.

Before proceeding on the analysis of the stochastic problem, it is
instructive to discuss briefly the deterministic limit $N \rightarrow \infty$%
. In this case the average number of master strings obeys the recursion
equation 
\begin{eqnarray}
\langle n \rangle_{t+1} & = & \sum_{n=0}^N \sum_{m=0}^N n \, T(n,m) \, {\cal 
{P}}_t (m)  \nonumber \\
& = & Q a ~\langle n \rangle_t ,
\end{eqnarray}
whose solution is $\langle n \rangle_t = \left ( Q a \right )^t \langle n
\rangle_0$. Hence, in the deterministic regime we find 
\begin{equation}  \label{t_det}
\tau = - \frac{1}{ \ln \left ( Q a \right )}
\end{equation}
which diverges at $Q = Q_c = 1/a$, signalling thus the existence of a phase
transition in the limit $N \rightarrow \infty$. Clearly, for $Q > Q_c$ the
master strings are always present in the population so that $\tau$ is
infinite in this entire region.

We consider now the finite $N$ regime. In this case the recursion equations
for the moments of $n$ do not yield useful information since, as usual, the
moment of order $p$ depends on the moment of order $p+1$ evaluated at the
previous generation. We resort then to a direct calculation of the
probability distribution ${\cal {P}}_t (m)$. More specifically, we will
focus on the calculation of ${\cal {P}}_t (0)$, since this is the quantity
that measures the rate of vanishing of the master strings from the
population. Although ${\cal {P}}_t (0)$ could be evaluated through a series
of matrix multiplications, a simple linear algebra calculation yields \cite
{Kimura} 
\begin{eqnarray}  \label{linear}
{\cal {P}}_t (0) & = & \sum_{n=0}^N c_{n} l_{n0} \lambda_n^t  \nonumber \\
& = & 1 + c_{1} l_{10} \lambda_1^t + \ldots + c_{N} l_{N0} \lambda_N^t
\end{eqnarray}
where $\lambda_n$ are the eigenvalues of ${\bf T}$, $l_{n0}$ are the zeroth
components of the eigenvectors ${\bf l}_n$, and $c_n$ are parameters that
depend on the initial state ${\cal {P}}_0 (n)$. Also we have used $\lambda_0
= l_{00} = c_0 = 1$. Assuming without loss of generality that $1 \geq
\lambda_1 \geq \ldots \geq \lambda_N \geq 0$, in the limit of large $t$ we
find 
\begin{equation}  \label{aprox}
1 - {\cal {P}}_t (0) \approx C \mbox{e}^{-t/\tau}
\end{equation}
where 
\begin{equation}  \label{tau}
\tau = - \frac{1}{ \ln \lambda_1 } ,
\end{equation}
and $C$ is a constant that depends on the initial state. Thus the problem
becomes the one of finding the second largest eigenvalue of the nonsymmetric
matrix ${\bf T}$. Since the largest eigenvalue and its corresponding
eigenvector are already known, this numerical problem yields easily to the
vector iteration method \cite{power}. Alternatively, we could find $\tau$ by
following the time evolution of ${\cal {P}}_t (0)$, obtained directly
through the recursion equations (\ref{recursion}), for a few generations and
then plotting $\ln \left [ 1 - {\cal {P}}_t (0) \right ]$ against the
generation number $t$. We have verified that both methods yield the same
results for $\tau$.

In Fig.\ 1 we present the dependence of $\ln \tau$ on the probability of
exact replication of an entire string, $Q$, for $a = 2$ and several values
of $N$. The finite $N$ effects are negligible for values of $Q$ smaller
than, though not too close, $Q_c$, as indicated by the very good agreement
between the finite $N$ data and the theoretical prediction for $N
\rightarrow \infty$ given in Eq.\ (\ref{t_det}). Since we expect $\tau$ to
increase exponentially with increasing $N$ for $Q > Q_c$, and to tend
towards its limiting value, Eq.\ (\ref{t_det}), also exponentially with $N$
for $Q < Q_c$, the issue is then to determine the dependence of $\tau$ on $N$
at the critical point $Q=Q_c$. In Fig.\ 2 we present $\ln \tau$ calculated
at $Q_c = 1/a$ against $\ln N$ for different values of $a$. These results
indicate clearly that at the critical point $\tau$ increases like $N^{1/2}$,
irrespective of the value of $a$. Once we have identified the rescaling of $%
\ln \tau$ that leads to the collapsing of the data for different $N$ at $%
Q=Q_c$, the next step is to determine the sharpness of the transition,
namely, the range of $Q$ about $Q_c$ where the transition characteristics
persists. This is achieved by assuming that the size of this region shrinks
to zero like $N^{-1/\nu}$ as $N \rightarrow \infty$, where the exponent $\nu
\geq 0$ is estimated using finite-size scaling or, more precisely, the data
collapsing method \cite{Binder}. In Fig.\ 3 we show the collapse of the data
for different $N$ obtained with $\nu = 2$ for $a= 2$, $10$ and $50$.
Although for $a=2$ we can achieve a good-quality data collapse using
relatively small population sizes ($N \geq 200$), for larger values of $a$,
however, a similar quality collapsing can only be obtained using larger
values of $N$ (i.e. $N \geq 400$). In summary, the results of the data
collapsing method indicate that the dependence of $\tau$ on $N$ in the
critical region is very well described by the scaling assumption 
\begin{equation}  \label{scaling}
\tau = N^{1/2} f_a \left [ \left ( Q - Q_c \right ) N^{1/2} \right ] ,
\end{equation}
where $f_a$ is a scaling function, whose specific form depends on the
parameter $a$.

To appreciate the effect of the selective advantage parameter $a$ on the
quality of the data collapsing results presented in Fig.\ 3, next we
consider in some detail the case $a \rightarrow \infty$ and $N$ finite.
Using $w_m \rightarrow 1$ for $m > 0$ yields 
\begin{equation}
T(n,m) = T(n) = \left ( \! \! 
\begin{array}{c}
N \\ 
n
\end{array}
\! \! \right ) \, Q^n \left (1 - Q \right )^{N-n} ~~~~~~~~m > 0.
\end{equation}
As before, $T(0,0) = 1$ and $T(n,0) = 0$ for $n > 0$. In this case the
eigenvalues of ${\bf T}$ can easily be calculated analytically yielding $%
\lambda_0 = 1$, $\lambda_1 = \sum_{n=1}^N T(n)$, $\lambda_2 = \ldots =
\lambda_N = 0$. Hence, 
\begin{equation}
\tau = - \frac{1}{ \ln \left [ 1 - \left ( 1 - Q \right )^N \right ] } \; .
\end{equation}
Finally, taking the limits $Q \rightarrow Q_c = 0$ and $N \rightarrow \infty$%
, we can easily verify that $\nu = 1$ in this limit. This interesting result
suggests that uncontrolled approximations and simplifications of the
original model which enhance the selective advantage of the master string or
the finite population sampling effects are expected to give unreliable
estimates of the exponent $\nu$. Moreover, care must be taken in restricting
the finite-size scaling analysis to the regime $N \gg a$ to avoid
underestimating the value of $\nu$. We note, of course, that the situation
of interest is $N \rightarrow \infty$ while $a$ remains finite.

To conclude, the collapse of the data for different $N$ into $a$-dependent
scaling functions presented in Fig.\ 3 and summarized in the scaling
assumption (\ref{scaling}) provide a full characterization of the error
threshold transition, signalled in our model by the divergence of $\tau$ at $%
Q_c = 1/a$, for large $N$. We emphasize that the main advantage of our
approach is that it does not rely upon any arbitrary definition of error
threshold for finite populations.

\bigskip

\parindent=0pt {\bf Acknowledgments} The work of JFF was supported in part
by Conselho Nacional de Desenvolvimento Cient\'{\i}fico e Tecnol\'ogico
(CNPq). PRAC is supported by FAPESP.

\newpage

\section*{Figure captions}

\bigskip

\parindent=0pt

{\bf Fig. 1} Logarithm of the characteristic time for the vanishing of the
master strings from the population, $\ln \tau$, as a function of the
probability of exact replication, $Q$, for $a = 2$, and $N = 100$ ($\ast$), $%
200$ ($\Diamond$), $300$ ($\Box$), $500$ ($\triangle$), and $600$ ($\times$%
). The solid line is the theoretical prediction for $N \rightarrow \infty$.

\bigskip

{\bf Fig. 2} Logarithm of the characteristic time for the vanishing of the
master strings from the population, $\ln \tau$, calculated at $Q_c = 1/a$ as
a function of the logarithm of the population size, $\ln N$, for $a = 2$ ($%
\bigcirc$), $10$ ($\triangle$), and $50$ ($\Box$).

\bigskip

{\bf Fig. 3} Properly rescaled logarithm of the characteristic time for the
vanishing of the master strings from the population, $\ln \tau / \ln N^{1/2}$%
, as a function of $\left ( Q - Q_c \right ) ~ N^{1/2}$ for (from top to
bottom at $Q=Q_c$) $a = 2$, $10$ and $50$. The convention is $N = 200$ ($%
\Diamond$), $300$ ($\Box$), $400$ ($\bigcirc$), $500$ ($\triangle$), and $600
$ ($\times$). For $a = 10$ and $50$ only the data for $N \geq 400$ are
presented.

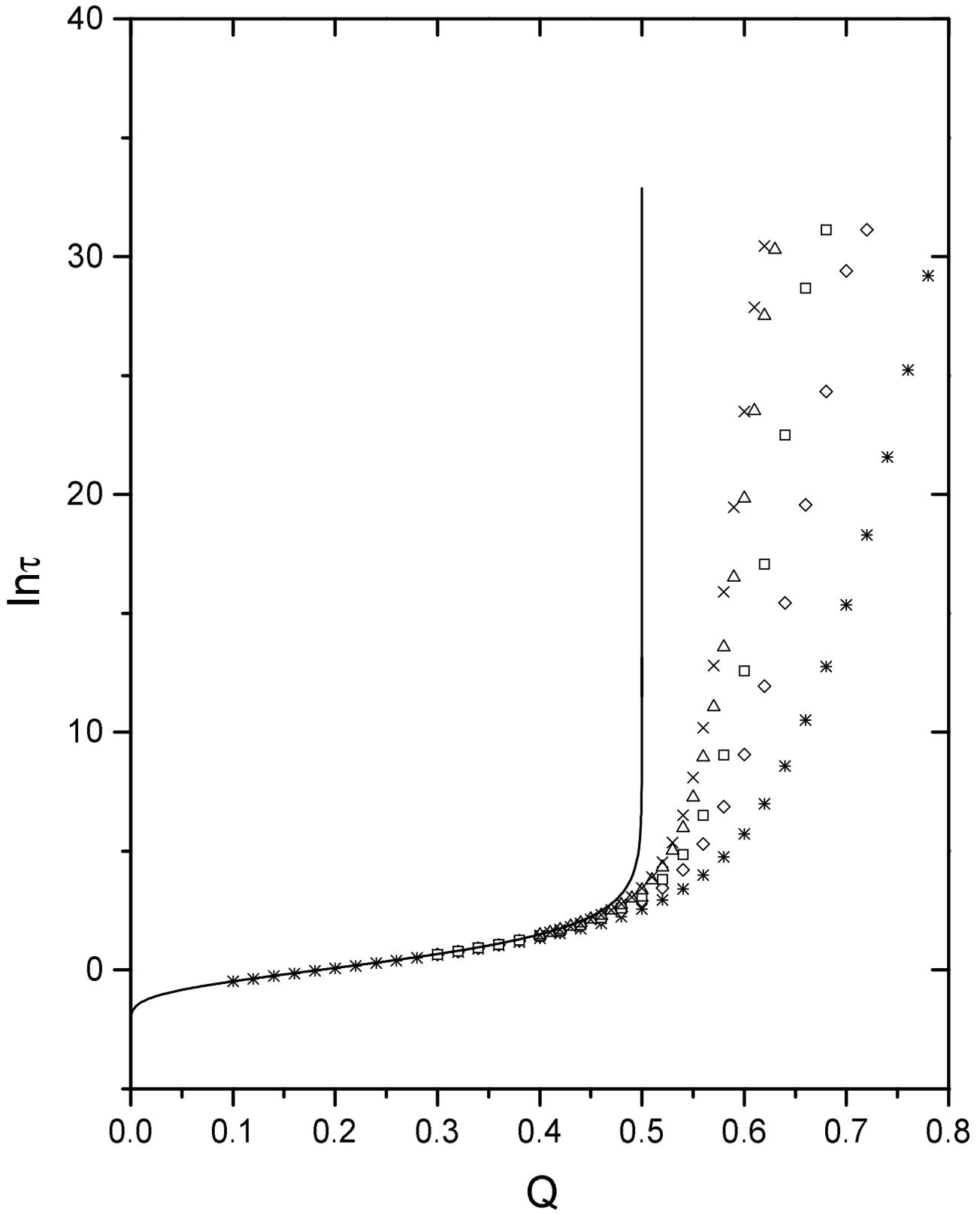
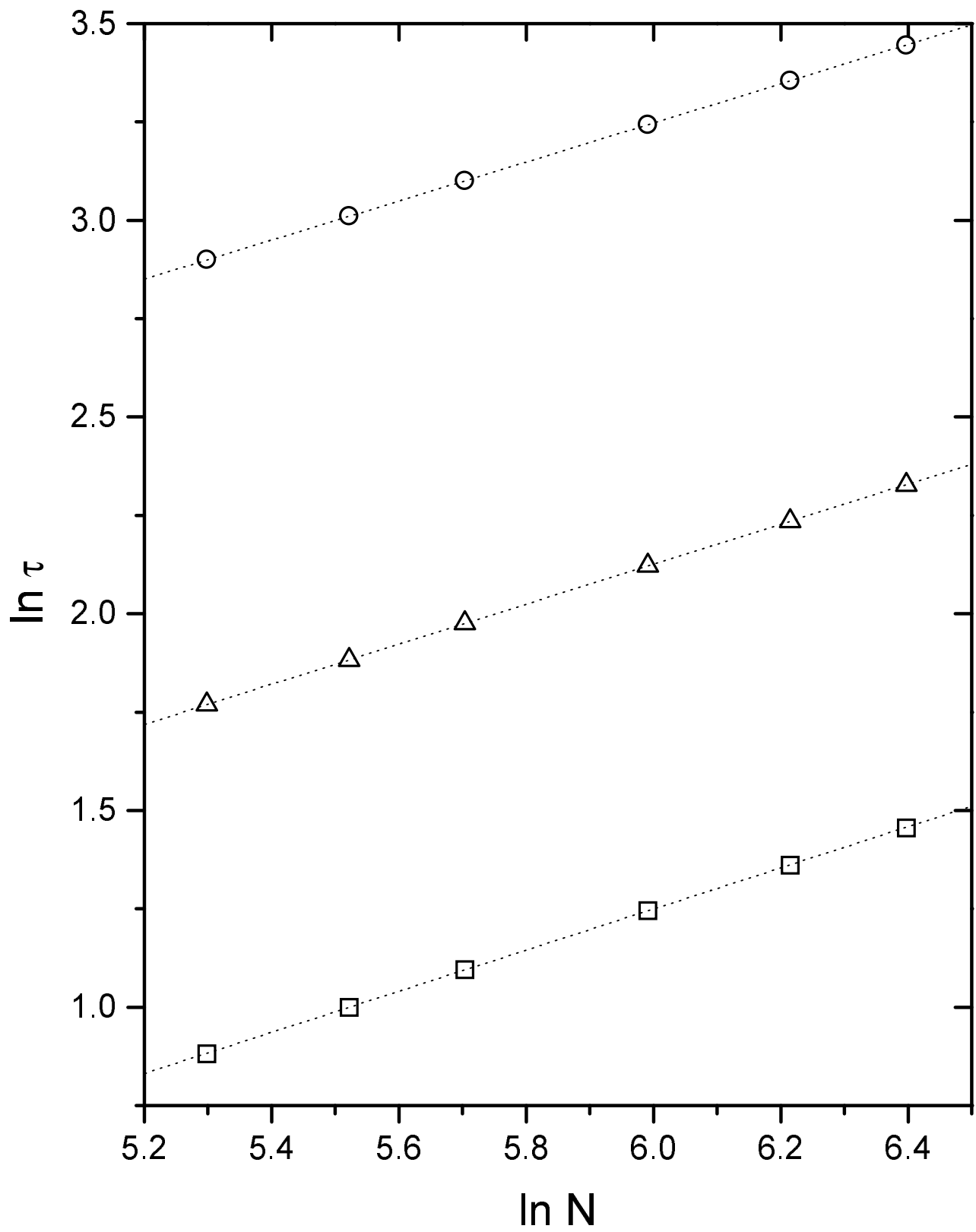
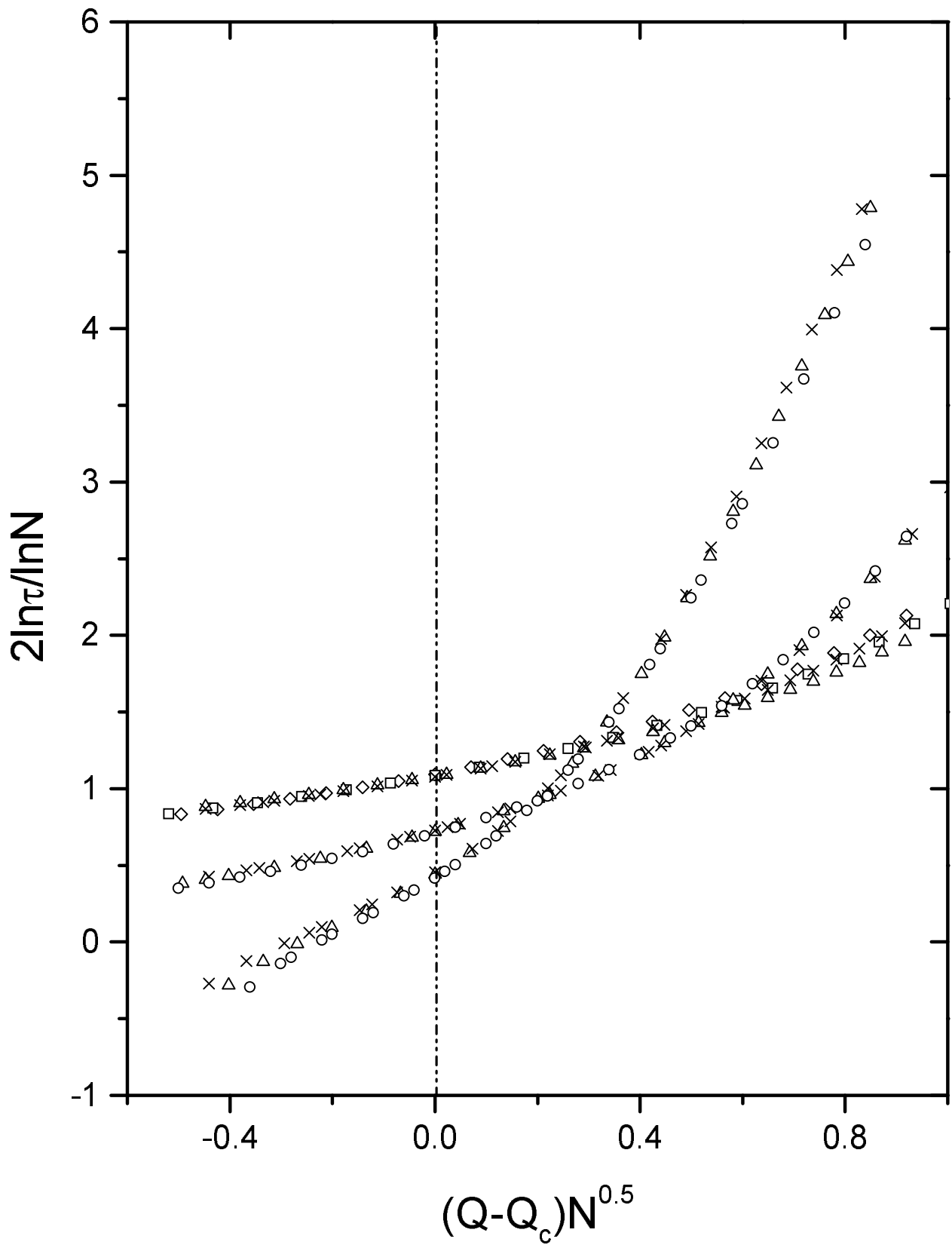
\end{document}